\documentclass[pra,showpacs,showkeys,byrevtex]{revtex4}%
\pdfoutput=1
\usepackage{amsfonts}
\usepackage{amsmath}
\usepackage{amssymb}
\usepackage{graphicx}%
\providecommand{\U}[1]{\protect\rule{.1in}{.1in}}
\newtheorem{theorem}{Theorem}

\newenvironment{proof}[1][Proof]{\noindent\textbf{#1.} }{\ \rule{0.5em}{0.5em}}
\begin{document}
\preprint{ }
\title[ ]{Identifying the quantum correlations in light-harvesting complexes}
\author{Kamil Br\'{a}dler}
\author{Mark M. Wilde}
\affiliation{School of Computer Science, McGill University, Montreal, Quebec, Canada H3A 2A7}
\author{Sai Vinjanampathy}
\affiliation{Department of Physics and Astronomy, Louisiana State University, Baton Rouge,
Louisiana, USA 70803}
\author{Dmitry B. Uskov}
\affiliation{Department of Physics and Engineering Physics, Tulane University, New Orleans,
Louisiana, USA 70118}
\keywords{light-harvesting complexes, quantum biology, quantum discord, relative entropy
of entanglement, quantum mutual information}
\pacs{03.67.Mn, 03.65.Yz, 82.39.Jn}

\begin{abstract}
One of the major efforts in the quantum biological program is to subject
biological systems to standard tests or measures of quantumness. These tests
and measures should elucidate if non-trivial quantum effects may be present in
biological systems. Two such measures of quantum correlations are the quantum
discord and the relative entropy of entanglement. Here, we show that the
relative entropy of entanglement admits a simple analytic form when dynamics
and accessible degrees of freedom are restricted to a zero- and
single-excitation subspace. We also simulate and calculate the amount of
quantum discord that is present in the Fenna-Matthews-Olson protein complex
during the transfer of an excitation from a chlorosome antenna to a reaction
center. We find that the single-excitation quantum discord and relative
entropy of entanglement are equal for all of our numerical simulations, but a
proof of their general equality for this setting evades us for now. Also, some
of our simulations demonstrate that the relative entropy of entanglement
without the single-excitation restriction is much lower than the quantum
discord. The first picosecond of dynamics is the relevant timescale for the
transfer of the excitation, according to some sources in the literature. Our
simulation results indicate that quantum correlations contribute a significant
fraction of the total correlation during this first picosecond in many cases,
at both cryogenic and physiological temperature.

\end{abstract}
\volumeyear{ }
\volumenumber{ }
\issuenumber{ }
\eid{ }
\date{\today}
\maketitle

\section{Introduction}

Quantum biology aims to understand if, how, and why biological systems exploit
quantum-mechanical effects for their functionality or for an evolutionary
advantage \cite{arndt:118,Lloyd:2009:164,AGDHZEWB08,ADP08book}. Exemplary
biological systems that may exploit quantum effects vary from photosynthetic
light-harvesting complexes \cite{CF09}, to the avian compass for bird
navigation \cite{RTPWW04}, to the olfactory system \cite{brookes:038101}.
Ongoing theoretical research indicates that light-harvesting complexes exploit
an environment-assisted quantum-walk like effect to enhance energy transport
\cite{mohseni:174106,1367-2630-11-3-033003,1367-2630-10-11-113019,caruso:105106}%
, the avian compass exploits a radical ion-pair mechanism for increased
sensitivity of the earth's geomagnetic field \cite{MHCKRLGTH08,K08c,CGB09},
and the olfactory system may exploit phonon-assisted tunneling for enhanced
detection of smell \cite{brookes:038101}.

Light-harvesting complexes seem particularly suitable as biological systems to
harness quantum-mechanical effects. Their lengthscales and energyscales are on
the order where we would expect quantum-mechanical laws to apply
\cite{Adolphs20062778}, but what remains less clear is if they can still
harness quantum effects such as entanglement even at physiological
temperature. A recent numerical study addresses this question by showing that
light-harvesting complexes could demonstrate stronger-than-classical
\textit{temporal }correlations, even at physiological temperature
\cite{WMM09}. However, it remains an open task to devise and conduct a
realistic experimental protocol that demonstrates an irrevocable test of
non-classical temporal correlations for light-harvesting complexes.

The aim of the study in Ref.~\cite{WMM09}, as well as that in
Refs.~\cite{SIFW09,CCDHP09}, is to address one of the important
\textquotedblleft quantum biological questions\textquotedblright%
:\ \textit{Does the biological system exhibit \textquotedblleft
quantumness\textquotedblright\ according to a standard test or measure?}
Theoretical machinery from quantum information science~\cite{horodecki:865},
developed specifically for understanding quantum computational and
communication devices, should be of immense utility in answering this
question. Several articles have already begun exploiting such tools.
Ref.~\cite{WMM09} exploits the Leggett-Garg test of non-classicality
\cite{LG85}\ to suggest that light-harvesting complexes might exhibit
stronger-than-classical temporal correlations, while
Refs.~\cite{SIFW09,CCDHP09} utilize standard entanglement-based measures of
quantum correlations \cite{HHH96,PhysRevLett.77.1413,PhysRevA.57.1619} to
suggest that they might exhibit stronger-than-classical spatial correlations.
Such standard measures of quantum behavior are more convincing than, say, a
claim that wavelike motion in population elements of a density matrix is a
signature of quantumness \cite{AkihitoIshizaki10132009}.

The studies in Refs.~\cite{SIFW09,CCDHP09} both suggest that it might be
possible to observe spatial quantum correlations in the Fenna-Matthews-Olson
(FMO)\ light-harvesting protein complex, but the quantum correlation measures
exploited there, such as concurrence~\cite{PhysRevLett.78.5022}, the measure
based on the global relative entropy of entanglement~\cite{PhysRevA.57.1619},
and logarithmic negativity~\cite{PhysRevLett.95.090503}, might not capture all
of the spatial quantum correlations that are present in a given quantum
system. For example, consider a bipartite quantum system in the state:%
\begin{equation}
\frac{1}{2}\left\vert 0\right\rangle \left\langle 0\right\vert ^{A}%
\otimes\left\vert +\right\rangle \left\langle +\right\vert ^{B}+\frac{1}%
{2}\left\vert -\right\rangle \left\langle -\right\vert ^{A}\otimes\left\vert
1\right\rangle \left\langle 1\right\vert ^{B}, \label{eq:sample-state}%
\end{equation}
where $\left\vert \pm\right\rangle \equiv\left(  \left\vert 0\right\rangle
\pm\left\vert 1\right\rangle \right)  /\sqrt{2}$ and Alice possesses the
system $A$ and Bob possesses the system $B$. The state in
(\ref{eq:sample-state}) is a separable state \cite{PhysRevA.40.4277}, meaning
that Alice and Bob can prepare it by means of local quantum operations and
classical communication. Yet, it is fundamentally non-classical because the
states $\left\vert 0\right\rangle $ and $\left\vert -\right\rangle $ on
Alice's local system or $\left\vert 1\right\rangle $ and $\left\vert
+\right\rangle $ on Bob's local system are indistinguishable from one another
(they are non-orthogonal). It could be possible for two bacteriochlorophyll
sites of a light-harvesting complex to admit a state of the above class or a
mixture of such states, but the quantum correlation measures mentioned above
all vanish for such a separable state, despite its status as a fundamentally
non-classical state. Thus, these measures might not capture the full
\textquotedblleft quantumness\textquotedblright\ that might be present in a
light-harvesting complex.

A different measure of quantum correlations, known as the quantum discord
\cite{0305-4470-34-35-315,PhysRevLett.88.017901}, captures all correlations in
a quantum state that are non-classical. The discord is a measure of the total
correlations present in a shared quantum state, reduced by the classical
correlations obtainable when one party performs a local measurement. For
example, the state in (\ref{eq:sample-state}) registers a non-vanishing
quantum discord and therefore possesses non-classical correlations according
to this measure, even though it does not violate a Bell inequality
\cite{B64}\ due to it being a separable state. Other examples demonstrate that
the quantum discord in a quantum system can remain positive even if the
entanglement vanishes after a finite time \cite{werlang:024103}. States that
register non-vanishing discord can lead to exponential speedups
\cite{datta:050502,lanyon:200501}\ in a quantum computational model known as
the \textquotedblleft one-clean qubit model\textquotedblright\ \cite{SJ08},
where the only requirement for a speedup is access to a single qubit in a pure
state. Although it is unlikely that a light-harvesting complex could be
performing exotic quantum computational speedups of the aforementioned nature
or in the standard way \cite{HSW09}, one cannot rule out the possibility that
a light-harvesting complex exhibiting non-vanishing discord could be
exploiting \textquotedblleft quantumness\textquotedblright\ of this form for
enhanced energy transport.

In this paper, we systematically study the presence of quantum correlations in
the FMO\ light-harvesting complex, both at cryogenic and physiological
temperature. Our first contribution is a simple formula for computing the
relative entropy of entanglement~\cite{RevModPhys.74.197} when a
light-harvesting complex evolves according to the open quantum system model
well studied in the quantum biological
literature~\cite{1367-2630-11-3-033003,1367-2630-10-11-113019}. This formula
assumes that dynamics are restricted to a zero- and single-excitation
subspace, and it reduces the \textit{a priori} computationally intensive
optimization task for computing the single-excitation relative entropy of
entanglement to a simple calculation with quantum entropies. These results for
the single-excitation relative entropy of entanglement generalize those of
Sarovar \textit{et al}.~in Ref.~\cite{SIFW09} for the global relative entropy
of entanglement. We then calculate the quantum discord for several
phenomenologically motivated \textquotedblleft bipartite
cuts\textquotedblright\ of the FMO protein complex and find that it is
equivalent to the single-excitation relative entropy of entanglement in all
cases presented here. These results demonstrate that quantum correlations
contribute a significant fraction of the total correlation during the highly
relevant first picosecond of dynamics in many cases (some sources in the
literature \cite{caruso:105106,1367-2630-10-11-113019}\ indicate that the
average time it takes for an excitation to trap to the reaction center is one
picosecond after it arrives from the chlorosome antenna). In other cases,
these quantities are equivalent and contribute a non-negligible fraction of
the total correlation during the first picosecond. This contribution indicates
that non-classical spatial correlations may be playing a role in the efficient
transfer of an excitation from the chlorosome antenna to the reaction center.
Our final contribution is to study the relative entropy of entanglement
without its optimization restricted to the single-excitation subspace, and we
find that it can be significantly less than the restricted relative entropy of
entanglement while only using just a small fraction of doubly-excited states
in the optimization.

We structure this paper as follows. We first briefly review both the dynamical
model of the FMO\ protein complex, the definition of the quantum discord, the
definition of the relative entropy of entanglement, and we then discuss how to
compute these quantities in the FMO\ protein complex model. Our analytic
result states that the relative entropy of entanglement, when restricted to
the zero- and single-excitation subspace, is equal to a simple formula that is
a difference of entropies. Appendix~\ref{sec:app-proof}\ provides a proof of
this theorem. Section~\ref{sec:results}\ presents the results of our numerical
simulations under various initial configurations and temperatures, and
Section~\ref{sec:unrestricted-REE}\ discusses the unrestricted optimization of
the relative entropy of entanglement. We then conclude with observations and
open questions for future research.

\section{Review}

\subsection{FMO Complex Dynamics}

The FMO\ protein complex is the crucial, light-harvesting component of the
green sulfur bacteria \textit{prosthecochloris aestuarii}, that develop in
dimly-lit, anoxic environments such as stratified lakes or sulfur springs
\cite{Fenna:1975:573}. It is a trimer, consisting of three identical subunits.
We study one unit of the trimer, consisting of seven bacteriochlorophyll sites
that act as a \textquotedblleft molecular wire,\textquotedblright%
\ transferring energetic excitations from a photon-receiving antenna to a
reaction center. A photon impinges on the antenna, producing an electronic
excitation, dubbed an exciton, that then proceeds to the unit with seven
sites. While traversing the seven sites, the exciton can either recombine,
corresponding to an energetic loss, or it can trap to the reaction center for
energy storage.

We characterize the quantum state of the exciton as a density operator in the
site basis:%
\[
\rho\equiv\sum_{m,n\in\left\{  G,1,\ldots,7,S\right\}  }\rho_{m,n}\left\vert
m\right\rangle \left\langle n\right\vert ,
\]
where the state $\left\vert m\right\rangle $ indicates that the exciton is
present at site $m$. The sites can be any of the seven sites in the protein
complex\ $m\in\left\{  1,\ldots,7\right\}  $, a ground state $\left\vert
G\right\rangle $ that represents the loss or recombination of the exciton, or
a sink state $\left\vert S\right\rangle $ that implies that the exciton has
trapped to the reaction center. We adopt the following \textquotedblleft
qubit\textquotedblright\ convention in this work (as adopted in previous works
\cite{SFM09,SIFW09,olaya-castro:085115}), by assigning site states to
tensor-product states:%
\begin{align*}
\left\vert G\right\rangle  &  \equiv\left\vert g\right\rangle _{1}\left\vert
g\right\rangle _{2}\left\vert g\right\rangle _{3}\left\vert g\right\rangle
_{4}\left\vert g\right\rangle _{5}\left\vert g\right\rangle _{6}\left\vert
g\right\rangle _{7}\left\vert g\right\rangle _{S},\\
\left\vert 1\right\rangle  &  \equiv\left\vert e\right\rangle _{1}\left\vert
g\right\rangle _{2}\left\vert g\right\rangle _{3}\left\vert g\right\rangle
_{4}\left\vert g\right\rangle _{5}\left\vert g\right\rangle _{6}\left\vert
g\right\rangle _{7}\left\vert g\right\rangle _{S},\\
&  \vdots\\
\left\vert 7\right\rangle  &  \equiv\left\vert g\right\rangle _{1}\left\vert
g\right\rangle _{2}\left\vert g\right\rangle _{3}\left\vert g\right\rangle
_{4}\left\vert g\right\rangle _{5}\left\vert g\right\rangle _{6}\left\vert
e\right\rangle _{7}\left\vert g\right\rangle _{S},\\
\left\vert S\right\rangle  &  \equiv\left\vert g\right\rangle _{1}\left\vert
g\right\rangle _{2}\left\vert g\right\rangle _{3}\left\vert g\right\rangle
_{4}\left\vert g\right\rangle _{5}\left\vert g\right\rangle _{6}\left\vert
g\right\rangle _{7}\left\vert e\right\rangle _{S},
\end{align*}
where $g$ indicates the absence of an excitation and $e$ indicates the
presence of an excitation at a particular site. The excitation number is a
conserved quantity in the absence of light-matter interaction events
\cite{TGOwens031987}, and this observation restricts the protein dynamics to a
zero- and single-excitation subspace. Thus, the above nine states are the only
states that we consider for the exciton while it traverses the seven sites.

Observe that \textquotedblleft tracing over any site\textquotedblright\ in
this qubit representation has the effect of placing the population term for
that site into the population of the ground state. For example, suppose that
we trace over all sites except for the first one. The resulting density matrix
has the form:%
\[
\left(  \rho_{GG}+\rho_{22}+\ldots+\rho_{77}+\rho_{SS}\right)  \left\vert
G\right\rangle \left\langle G\right\vert +\rho_{11}\left\vert 1\right\rangle
\left\langle 1\right\vert +\rho_{1G}\left\vert 1\right\rangle \left\langle
G\right\vert +\rho_{G1}\left\vert G\right\rangle \left\langle 1\right\vert .
\]
Such manipulations are important for computing correlation measures for any
bipartite cut of the sites in the FMO\ complex.

Evolution of the density matrix occurs according to a combination of both
coherent and incoherent dynamics. A tight-binding Hamiltonian of the following
form governs coherent evolution across the seven site states $\left\vert
1\right\rangle $, \ldots, $\left\vert 7\right\rangle $:%
\[
H\equiv\sum_{m}E_{m}\left\vert m\right\rangle \left\langle m\right\vert
+\sum_{n<m}V_{nm}\left(  \left\vert n\right\rangle \left\langle m\right\vert
+\left\vert m\right\rangle \left\langle n\right\vert \right)  ,
\]
where $E_{m}$ is the relative energy at site $m$ and $V_{nm}$ represents a
coupling between sites $n$ and $m$ (see
Refs.~\cite{Adolphs20062778,caruso:105106,WMM09}\ for the Hamiltonian that
governs evolution). The third site has the lowest energy of all seven sites
and is closest to the reaction center, and it is for these reasons that
researchers suggest that the objective of the molecular wire is to transmit
the excitation as quickly as possible to the third site
\cite{mohseni:174106,1367-2630-11-3-033003,1367-2630-10-11-113019,caruso:105106}%
. The off-diagonal terms lead to coherent couplings between sites---the
strongest couplings are between the first and second sites, the fourth and
fifth sites, and the fifth and sixth sites. Recent work determines the energy
landscape of the FMO\ complex that reveals coherent pathways the excitation
may traverse to get to the third site \cite{AkihitoIshizaki10132009,CDCHP09}.

A set of three Lindblad superoperators governs incoherent evolution of the
FMO\ complex, and each superoperator has a phenomenological justification for
its presence in the evolution
\cite{mohseni:174106,1367-2630-11-3-033003,caruso:105106,WMM09}. The first
incoherent mechanism $\mathcal{L}_{\text{diss}}$ is due to recombination or
loss of the exciton, acting like an amplitude damping mechanism. The second
incoherent mechanism $\mathcal{L}_{\text{sink}}$\ is due to trapping of the
excitation from the third site to the reaction center. The final decoherence
mechanism $\mathcal{L}_{\text{deph}}$ is a local dephasing at each site, due
to unavoidable interaction with the surrounding protein environment.
Ref.~\cite{1367-2630-11-3-033003}\ shows how to relate the dephasing rate to
the temperature of the system, under a particular decoherence model (we employ
this relation in our numerical study in this paper).
Refs.~\cite{mohseni:174106,1367-2630-11-3-033003,caruso:105106,WMM09}%
\ describe in detail the decoherence mechanisms in the FMO\ complex. Despite
the Markovian nature of the evolution, it describes the dynamics reasonably
accurately \cite{1367-2630-11-3-033003} (though see the recent work on
non-Markovian evolution in the FMO\ complex \cite{RCA09,CCDHP09}).

Putting everything together, the following master equation gives the full
evolution of the density matrix $\rho$:%
\begin{equation}
\dot{\rho}=-i\left[  H,\rho\right]  +\mathcal{L}_{\text{diss}}\left(
\rho\right)  +\mathcal{L}_{\text{sink}}\left(  \rho\right)  +\mathcal{L}%
_{\text{deph}}\left(  \rho\right)  . \label{eq:evolution}%
\end{equation}
Note that the above evolution induces coherences only between sites in the
FMO\ complex, so that any off-diagonal element of the density matrix of the
form $\rho_{i,G}$ or $\rho_{i,S}$, where $i\in\left\{  1,\ldots,7\right\}  $,
vanishes. Various sources in the literature \cite{1367-2630-11-3-033003} and
recent experimental evidence \cite{JianzhongWen04142009} suggest that the
initial state of the FMO\ complex is a pure excitation at site one, a pure
excitation at site six, or a mixture of these two states because sites one and
six are closest to the antenna. We incorporate each of these cases in our simulations.

The dynamical model in (\ref{eq:evolution}) implies that the FMO\ complex
exhibits both classical and quantum random walk behavior, but does not exhibit
exotic \textquotedblleft quantum stochastic walk\textquotedblright\ behavior
according to Table~I of Ref.~\cite{RWA09}. The classical and quantum random
walk behavior then suggests that this evolution should establish both
classical and quantum correlation between sites in the complex. It is these
correlations that we measure in this paper.

\subsection{Quantum Discord}

We first recall various informational measures of a quantum state before
briefly reviewing the motivation for the quantum discord as a measure of
quantum correlations. Suppose that two parties Alice and Bob share a quantum
state $\rho^{AB}$. The von Neumann entropy of this state is as follows:%
\[
H\left(  AB\right)  _{\rho}\equiv-\text{tr}\left\{  \rho^{AB}\log\rho
^{AB}\right\}  ,
\]
where the logarithm is base two. The entropy $H\left(  AB\right)  _{\rho}$
measures the uncertainty about the total quantum state $\rho^{AB}$ in units of
bits, whenever the logarithm is base two \cite{PhysRevA.51.2738}. Similarly,
we can compute the marginal entropies $H\left(  A\right)  $ and $H\left(
B\right)  $---these entropies are with respect to the respective reduced
states $\rho^{A}$ and $\rho^{B}$, obtained by a respective partial trace over
Bob or Alice's share of the state $\rho^{AB}$.

The quantum mutual information $I\left(  A;B\right)  $ is a measure of the
total correlations, both classical and quantum, shared between two parties,
where%
\begin{equation}
I\left(  A;B\right)  _{\rho}\equiv H\left(  A\right)  _{\rho}+H\left(
B\right)  _{\rho}-H\left(  AB\right)  _{\rho}. \label{eq:MI}%
\end{equation}
The quantum mutual information admits a natural operational interpretation as
the amount of noise necessary to destroy all correlations present in a
bipartite state \cite{PhysRevA.72.032317}. Its other operational
interpretations are as the maximum amount of secret information that one party
can send to another if they use a shared state as the basis for a one-time pad
cryptosystem \cite{schumacher:042305}, the entanglement-assisted classical
capacity of a quantum channel \cite{PhysRevLett.83.3081,BSST01,arx2004shor},
and more recently the amount of private quantum information that a sender can
transmit to a receiver while being eavesdropped on by a uniformly accelerating
third party \cite{1126-6708-2009-08-074}. All of these operational
interpretations justify the notion of the quantum mutual information measuring
correlations between two parties in units of bits.

Suppose now that Alice would like to extract classical information from the
quantum state $\rho^{AB}$. She performs a von Neumann measurement $\left\{
\left\vert x\right\rangle \left\langle x\right\vert \right\}  $ on her share
of the state, where the states $\left\{  \left\vert x\right\rangle \right\}  $
form an orthonormal basis. If Alice obtains classical result $x$ from the
measurement, the resulting conditional quantum state of the whole system is
\[
\left\vert x\right\rangle \left\langle x\right\vert ^{X}\otimes\rho_{x}^{B},
\]
where%
\begin{gather}
\rho_{x}^{B}\equiv\frac{1}{p\left(  x\right)  }\text{Tr}_{X}\left\{
(\left\vert x\right\rangle \left\langle x\right\vert ^{X}\otimes I^{B}%
)\rho^{AB}(\left\vert x\right\rangle \left\langle x\right\vert ^{X}\otimes
I^{B})\right\}  ,\label{eq:cond-density-ops}\\
p\left(  x\right)  \equiv\text{Tr}\left\{  \left\vert x\right\rangle
\left\langle x\right\vert ^{X}\rho^{A}\right\}  , \label{eq:class-probs}%
\end{gather}
and we now label Alice's system with $X$ because it is classical. If we then
take the expectation over all of Alice's outcomes, the description of the
system is a classical-quantum state:%
\begin{equation}
\sigma^{XB}\equiv\sum_{x}p\left(  x\right)  \left\vert x\right\rangle
\left\langle x\right\vert ^{X}\otimes\rho_{x}^{B}. \label{eq:cq-state}%
\end{equation}
We might think that a natural measure of the classical correlations present in
a bipartite state is the amount of correlations in the resulting
classical-quantum state:%
\begin{equation}
I_{c}\left(  A;B\right)  _{\rho}\equiv\max_{\left\{  \Lambda^{A}\right\}
}I\left(  X;B\right)  _{\sigma}, \label{eq:classical-corr}%
\end{equation}
The above expression features a maximization over all of the von Neumann
measurements $\Lambda^{A}$\ that Alice could perform. These measurements
result in a state of the form $\sigma$ in (\ref{eq:cq-state}), and the quantum
mutual information $I\left(  X;B\right)  _{\sigma}$ is with respect to such a
state. In fact, Henderson and Vedral proposed such a measure
\cite{0305-4470-34-35-315}, and Devetak and Winter later justified this
measure by providing an operational interpretation of it as the maximum amount
of common randomness (perfect classical correlations) that two parties can
extract from a bipartite quantum state \cite{DW03a}.

The two correlation measures in (\ref{eq:MI}) and (\ref{eq:classical-corr}),
the first a measure of total correlations and the second a measure of
classical correlations, suggests that a measure of the quantum correlations
should be the difference of these two quantities. The \textit{quantum discord}
$D\left(  A;B\right)  _{\rho}$ is such a measure
\cite{PhysRevLett.88.017901,D08}, defined as the difference of total and
classical correlations:%
\begin{equation}
D\left(  A;B\right)  _{\rho}\equiv I\left(  A;B\right)  _{\rho}-I_{c}\left(
A;B\right)  _{\rho}. \label{eq:discord-horo}%
\end{equation}
The discord is always non-negative, by the quantum data processing inequality
of quantum information theory \cite{Nielsen:2000:CambridgeUniversityPress},
and it is generally not symmetric with respect to the parties $A$ and $B$:%
\[
\exists\rho:D\left(  A;B\right)  _{\rho}\neq D\left(  B;A\right)  _{\rho}.
\]
It captures all of the quantum correlations, including entanglement, but also
the quantumness in separable states of the form in (\ref{eq:sample-state}).
Zurek has suggested a physical interpretation of the discord as the difference
in efficiency between quantum and classical Maxwell's demons
\cite{PhysRevA.67.012320}, but it still lacks a clear operational
interpretation in the sense of
Refs.~\cite{PhysRevA.72.032317,schumacher:042305,PhysRevLett.83.3081,BSST01,arx2004shor,1126-6708-2009-08-074,DW03a}%
. Nevertheless, we still employ the discord as a measure of the quantum
correlations in the FMO\ complex.

We can rewrite the quantum discord as the following expression:%
\begin{equation}
D\left(  A;B\right)  _{\rho}=I\left(  B\rangle A\right)  _{\rho}%
+\min_{\left\{  \Pi_{x}\right\}  }\sum_{x}p\left(  x\right)  H\left(  \rho
_{x}^{B}\right)  , \label{eq:discord-formula}%
\end{equation}
by performing straightforward manipulations of the entropies in
(\ref{eq:discord-horo}). In the above, $I\left(  A\rangle B\right)  $ is the
coherent information~\cite{Nielsen:2000:CambridgeUniversityPress}, equal to
the negative of a conditional entropy:%
\[
I\left(  B\rangle A\right)  _{\rho}=-H\left(  B|A\right)  _{\rho}=H\left(
A\right)  _{\rho}-H\left(  AB\right)  _{\rho},
\]
and the probabilities $p\left(  x\right)  $ and conditional density operators
are as they appear respectively in (\ref{eq:cond-density-ops}) and
(\ref{eq:class-probs}).

\subsection{Relative Entropy of Entanglement}

The relative entropy of entanglement is an entanglement measure from quantum
information theory~\cite{RevModPhys.74.197}, and we briefly review its
definition. The relative entropy $D\left(  \rho^{AB}||\omega^{AB}\right)
$\ of two bipartite states $\rho^{AB}$ and $\omega^{AB}$ is as
follows~\cite{Nielsen:2000:CambridgeUniversityPress}:%
\begin{equation}
D\left(  \rho^{AB}||\omega^{AB}\right)  \equiv\text{Tr}\left\{  \rho^{AB}%
\log\rho^{AB}\right\}  -\text{Tr}\left\{  \rho^{AB}\log\omega^{AB}\right\}
.\label{eq:REE}%
\end{equation}
This measure, in some sense, quantifies the \textquotedblleft
distance\textquotedblright\ between two bipartite states, but it is not a
distance measure in the strict mathematical sense because it fails to be
symmetric. Though, this intuition is useful for constructing an entanglement
measure. We might naturally expect a good measure of entanglement to be the
distance of a given bipartite state to the closest separable state
$\sigma^{AB}\equiv\sum_{i}p\left(  i\right)  \sigma_{i}^{A}\otimes\sigma
_{i}^{B}$. The relative entropy of entanglement $R\left(  \rho^{AB}\right)
$\ is such a measure, defined as the minimization of the relative entropy over
all separable states:%
\[
R\left(  \rho^{AB}\right)  \equiv\min_{\sigma\in\mathcal{S}}D\left(  \rho
^{AB}||\sigma^{AB}\right)  ,
\]
where $\mathcal{S}$ is the class of separable states. This measure satisfies
the properties of an entanglement measure and appears extensively in the
quantum information theory literature~\cite{RevModPhys.74.197}. We mention
that Ref.~\cite{SIFW09}\ employed this entanglement measure for determining
quantum correlations in the FMO\ complex, but they considered the global
relative entropy of entanglement rather than a bipartite version of it as we
consider in this work.

\section{Formula for the Single-Excitation Relative Entropy of Entanglement}

The restriction of dynamics to the zero- and single-excitation subspace allows
for significant simplifications to the theory of energetic transfer in the
FMO\ complex. We can also apply this restriction to the relative entropy of
entanglement, by restricting the optimization over separable states to the
zero- and single-excitation subspace. Let $R_{e}\left(  \rho^{AB}\right)  $
denote the relative entropy of entanglement with this restriction applied to
its optimization. Theorem~\ref{thm:main-theorem}\ below states that
$R_{e}\left(  \rho^{AB}\right)  $ is equal to a simple formula that is a
difference of entropies. Thus, the theorem significantly simplifies the
computation of this quantity.

\begin{theorem}
\label{thm:main-theorem}Consider a density operator $\rho^{AB}$ restricted to
the zero- and single-excitation subspace. Let%
\[
\overline{\Delta}\left(  \rho^{AB}\right)  \equiv\alpha\left\vert
G\right\rangle \left\langle G\right\vert +\rho_{e}^{A}\otimes|G\rangle\langle
G|^{B}+|G\rangle\langle G|^{A}\otimes\rho_{e}^{B},
\]
where $\alpha$ is the population of the ground state, $\rho_{e}^{A}$ is the
projection of Alice's part of $\rho^{AB}$ into the single-excitation subspace,
and $\rho_{e}^{B}$ is defined in a similar way. Suppose that the dynamics for
this density operator never induce any coherences between the zero- and
single-excitation subspaces (as in the dynamics in (\ref{eq:evolution})). Then
the single-excitation relative entropy of entanglement $R_{e}\left(
\rho\right)  $ is equal to the difference of the entropy of $\overline{\Delta
}\left(  \rho^{AB}\right)  $ and $\rho^{AB}$:%
\[
R_{e}\left(  \rho^{AB}\right)  =H\left(  \overline{\Delta}\left(  \rho
^{AB}\right)  \right)  -H\left(  \rho^{AB}\right)  .
\]

\end{theorem}

The full proof appears in Appendix~\ref{sec:app-proof}. It exploits standard
properties of the von Neumann entropy and a perturbative argument.

\section{Simulation Results}

\label{sec:results}We conducted several simulations at both cryogenic
temperature (77$%
{{}^\circ}%
$K)\ and physiological temperature (300$%
{{}^\circ}%
$K) and for the initial state being a pure state at site one, six, and the
mixture of the previous two. These simulations calculate both the quantum
mutual information, quantum discord, and single-excitation relative entropy of
entanglement with respect to several \textquotedblleft bipartite
cuts\textquotedblright\ of the sites in the FMO\ complex. Throughout this
section, we refer to quantum discord, but all our simulations indicated that
the single-excitation relative entropy of entanglement is equal to the quantum
discord. These results provide strong evidence that the quantum discord is
equal to the formula in Theorem~\ref{thm:main-theorem}\ for these cases, but a
general proof of this conjecture eludes us for now. Our different simulation
cases are as follows:

\begin{enumerate}
\item We first considered the cut where system $A$ consists of site three and
system $B$ consists of sites one and six. We picked this cut because the
initial state of the complex is at site one, six, or the mixture, and the
objective of the FMO\ \textquotedblleft molecular wire\textquotedblright\ is
to transfer the excitation from these initial sites to site three. If spatial
quantum correlations play a role in the transfer of the excitation, one would
expect a state with this bipartite cut to register a non-negligible amount of
quantum discord.

\item The next bipartite cut that we considered is with the $A$ system
consisting of sites one and two and the $B$ system consisting of site three.
We picked such a cut because recent analysis of the FMO\ Hamiltonian
\cite{CDCHP09}\ suggests that a superposition state of sites one and two gives
an efficient energetic pathway for the excitation to transfer to site three.
The energy of the superposed state is closer to the energy of site three than
it is to the energy of either site one or site two. We would again expect such
a cut would register a non-negligible amount of quantum discord if quantum
correlations play a role in the transfer of the excitation.

\item Finally, after conducting the above simulations, we conducted
simulations with respect to the cut where system $A$ consists of site three
and system $B$ consists of all other sites. The quantum mutual information for
this case should be larger than for the case of the other cuts, by the quantum
data processing inequality.
\end{enumerate}

\begin{figure}
[ptb]
\begin{center}
\includegraphics[
natheight=4.786700in,
natwidth=17.012600in,
height=2.0046in,
width=7.0534in
]%
{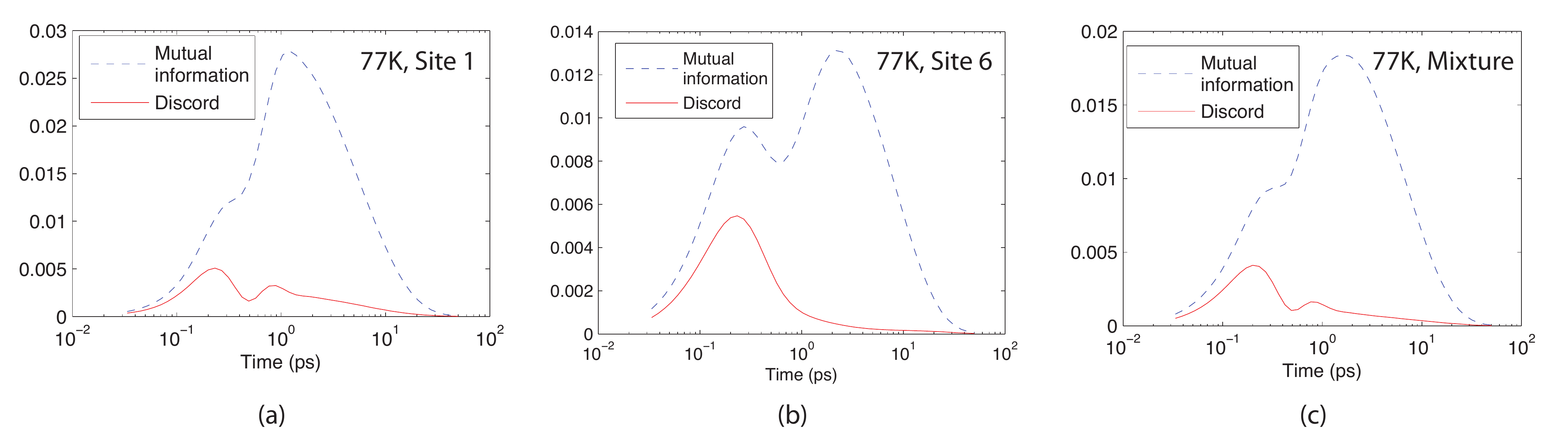}%
\caption{(Color online) Mutual information $I\left(  A;B\right)  $ and quantum
discord $D\left(  A;B\right)  $ with system $A$ as site three and system $B$
as sites one and six. The figure plots these quantities at cryogenic
temperature (77${{}^\circ}$K) as a function of time when the initial state is
(a) a pure state at the first site, (b) a pure state at the sixth state, (c)
an equal mixture of the two previous states. In each of the above plots, the
quantum discord is a significant fraction of the total correlations during the
first picosecond.}%
\label{fig:77K-MI-discord}%
\end{center}
\end{figure}

Figure~\ref{fig:77K-MI-discord}\ plots the results of the first simulation for
cryogenic temperature and for each initial state. The figure indicates that
the quantum discord contributes a significant fraction of the total
correlation for short timescales (less than one picosecond). This contribution
of the quantum discord to total correlation is significant, considering that
the average transfer time of the excitation to site three occurs around the
order of one picosecond in our model. For longer timescales (greater than one
picosecond), the total correlation between $A$ (sites one and six) and $B$
(site three) increases to its maximum at around 2-4 ps, and the quantum
discord no longer contributes any significant amount to the total correlation.
Even though the total correlation rises so much higher for longer timescales
than it is for shorter timescales, the increase and peak are not relevant for
excitation transfer, i.e., \textquotedblleft they arrive too
late\textquotedblright\ given that this transfer occurs on the order of 1 ps.
The total correlation then washes away as time increases beyond 10 ps, and it
should nearly vanish for times around 1 ns because this time is the average
recombination time of the exciton, and no correlations should persist after a
recombination.%
\begin{figure}
[ptb]
\begin{center}
\includegraphics[
natheight=4.739200in,
natwidth=16.726299in,
height=2.0185in,
width=7.0534in
]%
{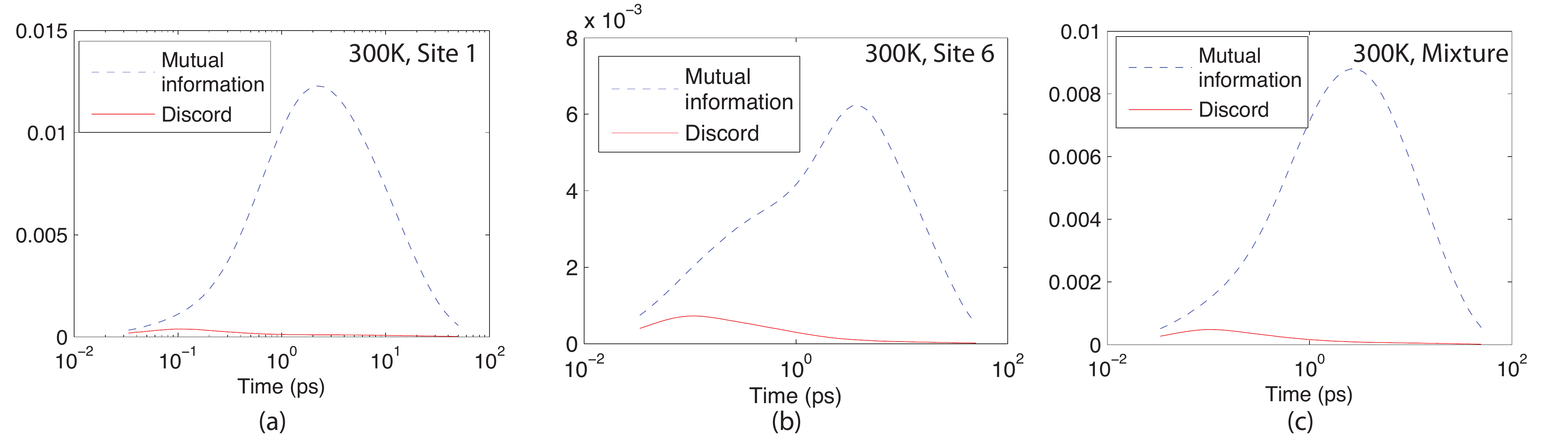}%
\caption{(Color online) Mutual information $I\left(  A;B\right)  $ and quantum
discord $D\left(  A;B\right)  $ with system $A$ as site three and system $B$
as sites one and six. The figure plots these quantities at physiological
temperature (300${{}^\circ}$K) as a function of time when the initial state is
(a) a pure state at the first site, (b) a pure state at the sixth state, (c)
an equal mixture of the two previous states. In each of the above figures, the
quantum discord is a non-negligible fraction of the total correlation during
the first picosecond.}%
\label{fig:MI-discord-300K}%
\end{center}
\end{figure}
%

\begin{figure}
[ptb]
\begin{center}
\includegraphics[
natheight=3.000000in,
natwidth=10.199600in,
height=2.0937in,
width=7.0534in
]%
{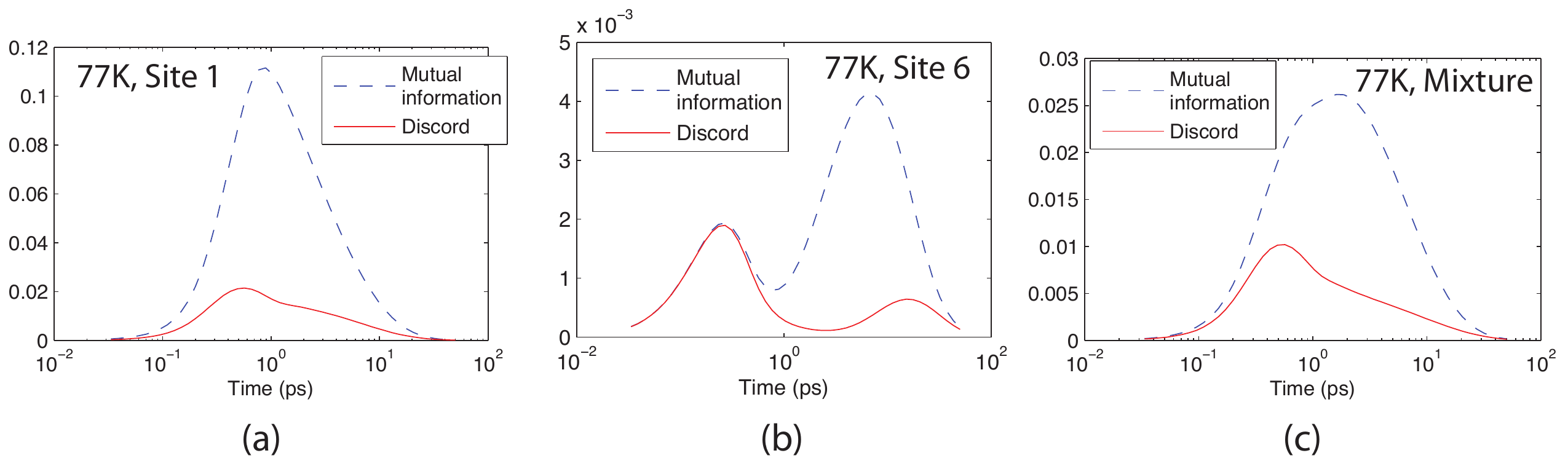}%
\caption{(Color online) Mutual information $I\left(  A;B\right)  $ and quantum
discord $D\left(  A;B\right)  $ with system $A$ as site three and system $B$
as sites one and two. The figure plots these quantities at cryogenic
temperature (77${{}^\circ}$K) as a function of time when the initial state is
(a) a pure state at the first site, (b) a pure state at the sixth state, (c)
an equal mixture of the two previous states. In each of the above cases, the
quantum discord is a significant fraction of the total correlation during the
first picosecond. In fact, for the second case, the quantum discord
contributes nearly all of the total correlation during the first picosecond.}%
\label{fig:77K-12-vs-3}%
\end{center}
\end{figure}
\begin{figure}
[ptb]
\begin{center}
\includegraphics[
natheight=2.987100in,
natwidth=10.120000in,
height=2.1006in,
width=7.0534in
]%
{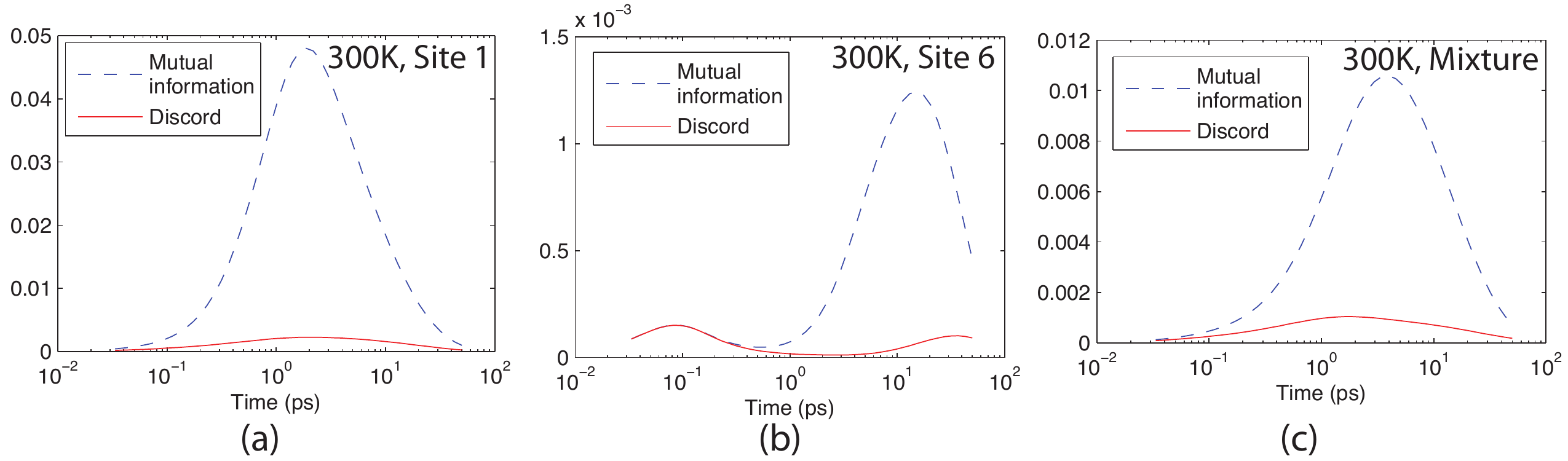}%
\caption{(Color online) Mutual information $I\left(  A;B\right)  $ and quantum
discord $D\left(  A;B\right)  $ with system $A$ as site three and system $B$
as sites one and two. The figure plots these quantities at physiological
temperature (300${{}^\circ}$K) as a function of time when the initial state is
(a) a pure state at the first site, (b) a pure state at the sixth state, (c)
an equal mixture of the two previous states. In each of the above cases, the
quantum discord contributes a fraction of the total correlation during the
first picosecond. In the second case, it again contributes nearly all of the
total correlation during the first picosecond.}%
\label{fig:300K-12-vs-3}%
\end{center}
\end{figure}

Figure~\ref{fig:MI-discord-300K}\ plots the results of the first simulation
for physiological temperature. The difference between cryogenic and
physiological temperature is a qualitative shrinking by a factor of two, but
the fraction of quantum discord that contributes to total correlation for
short timescales is lower than it is for cryogenic temperatures. The local
dephasing at each site acts to destroy both total and quantum correlation, but
it appears to have a more harmful effect on quantum correlation. Despite the
low amount of quantum discord registered at physiological temperature, it
could still be that a light-harvesting complex is harnessing this small amount
of quantumness (that is a significant fraction of total correlation) on this
short timescale to enhance excitation transfer.

Figures~\ref{fig:77K-12-vs-3} and \ref{fig:300K-12-vs-3}\ plot the results of
the second simulation for both respective temperatures and for each initial
state. These results are qualitatively similar to the previous bipartite cut,
but the most striking difference is that the quantum discord contributes all
of the correlation (it is equal to the quantum mutual information) for short
timescales when the initial state is at site six for both cryogenic and
physiological temperatures (see Figures~\ref{fig:77K-12-vs-3}(b) and
\ref{fig:300K-12-vs-3}(b)).

\begin{figure}
[ptb]
\begin{center}
\includegraphics[
natheight=2.853000in,
natwidth=10.326700in,
height=1.97in,
width=7.0534in
]%
{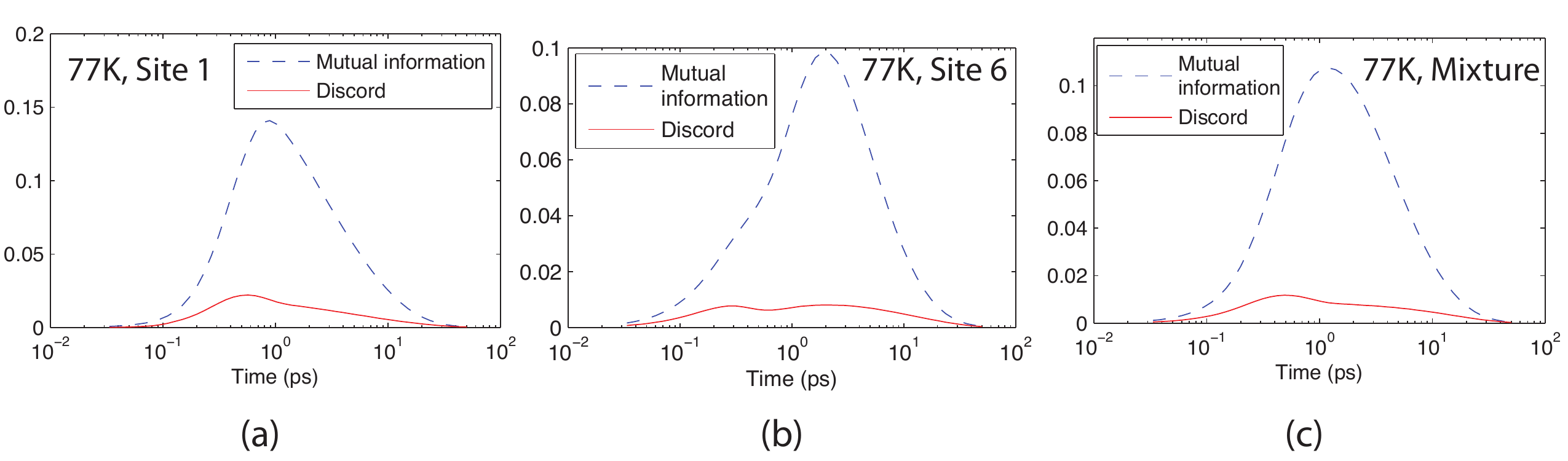}%
\caption{(Color online) Mutual information $I\left(  A;B\right)  $ and quantum
discord $D\left(  A;B\right)  $ with system $A$ as site three and system $B$
as all other sites (1-2, 4-7). The figure plots these quantities at cryogenic
temperature (77${{}^\circ}$K) as a function of time when the initial state is
(a) a pure state at the first site, (b) a pure state at the sixth state, (c)
an equal mixture of the two previous states. The quantum discord contributes a
significant fraction of the total correlation during the first picosecond.}%
\label{fig:77K-all-vs-3}%
\end{center}
\end{figure}

\begin{figure}
[ptb]
\begin{center}
\includegraphics[
natheight=2.987100in,
natwidth=9.986900in,
height=2.1283in,
width=7.0534in
]%
{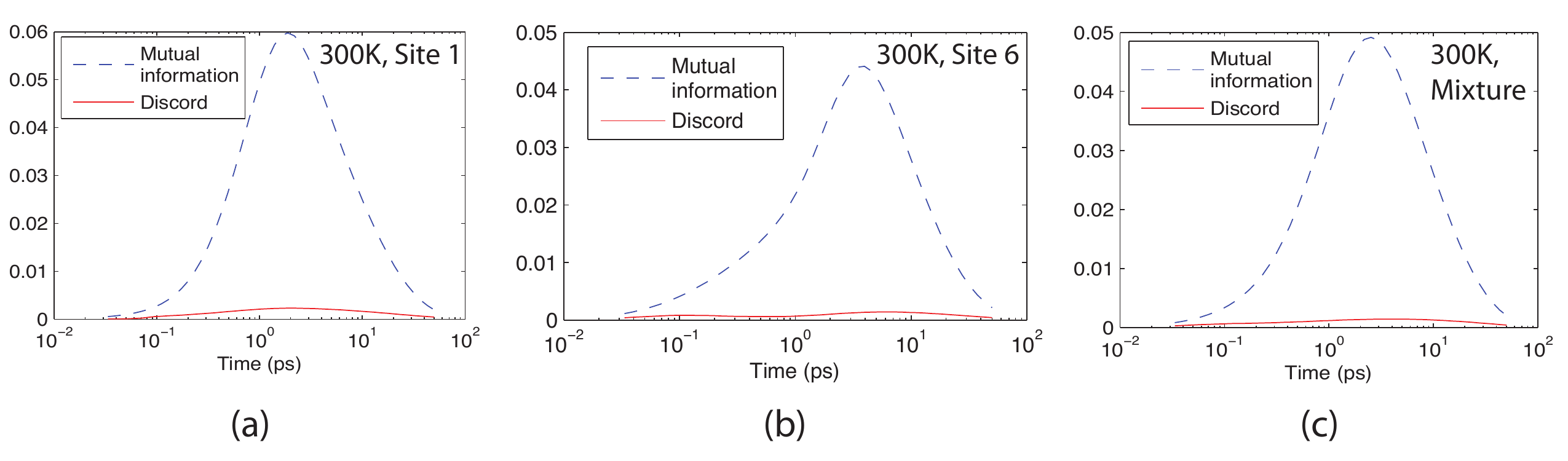}%
\caption{(Color online) Mutual information $I\left(  A;B\right)  $ and quantum
discord $D\left(  A;B\right)  $ with system $A$ as site three and system $B$
as all other sites (1-2, 4-7). The figure plots these quantities at
physiological temperature (300${{}^\circ}$K) as a function of time when the
initial state is (a) a pure state at the first site, (b) a pure state at the
sixth state, (c) an equal mixture of the two previous states. The quantum
discord does not contribute a large fraction of the total correlation here.}%
\label{fig:300K-all-vs-3}%
\end{center}
\end{figure}

Figures~\ref{fig:77K-all-vs-3} and \ref{fig:300K-all-vs-3}\ plot the results
of our final simulation, where the bipartite cut has system $B$ as site three
and system $A$ as all other sites. The results for quantum discord are again
qualitatively similar to previous results, but the quantum mutual information
is significantly higher than for the previous cases (this is expected because
the quantum data processing inequality states that correlations can only
decrease when discarding subsystems). The quantum discord again contributes a
significant fraction of the total correlation for short timescales and
contributes only a small fraction for later timescales. Again, physiological
temperature decoherence mitigates the presence of quantumness.

\section{Comparison of the Quantum Discord with the Standard Relative Entropy
of Entanglement}

\label{sec:unrestricted-REE}In this section, we focus on computing the
relative entropy of entanglement in (\ref{eq:REE}) without the assumption that
the set of separable states $\sigma^{AB}$ involved in the minimization of
$D(\rho^{AB}||\sigma^{AB})$ is limited to the single-excitation subspace. This
implies that $\sigma^{AB}$ may include multiply-excited states (for example,
states for which both systems $A$ and $B$ carry an exciton).

\begin{figure}
[ptb]
\begin{center}
\includegraphics[
natheight=10.026600in,
natwidth=22.893299in,
height=2.885in,
width=6.5518in
]%
{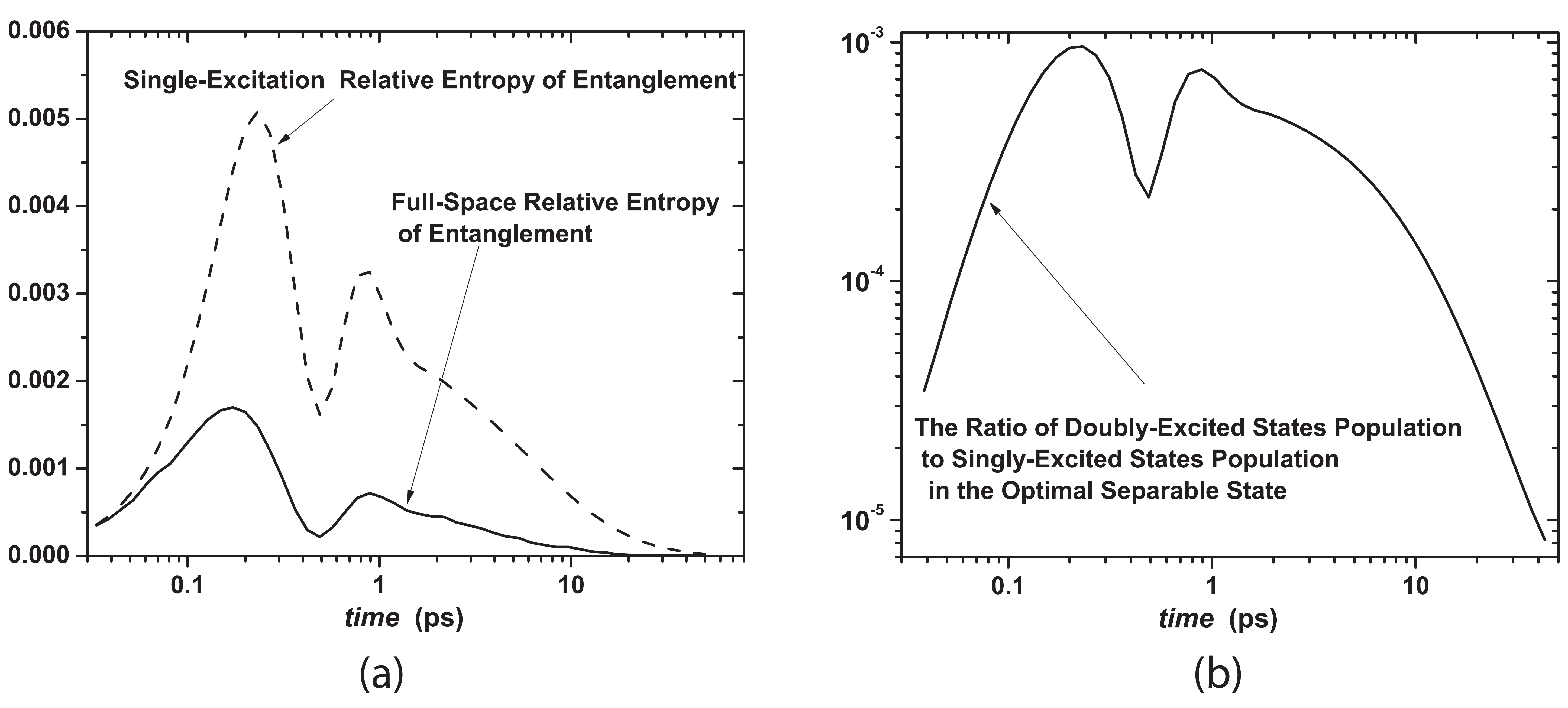}%
\caption{(a) This figure simulates the dynamics of the FMO\ complex at
77${{}^\circ}$K and calculates both the single-excitation relative entropy of
entanglement and the full relative entropy of entanglement for $A=1,6$ and
$B=3$. (b) The ratio of the number of doubly-excited states to the number of
singly-excited states in the optimal separable state for the relative entropy
of entanglement is rather small.}%
\label{fig:dima-plots}%
\end{center}
\end{figure}

A question may arise as to why such multiply-excited states are relevant for
the optimization of the relative entropy of entanglement. After all, the
Hamiltonian of the system commutes with the operator corresponding to the
total number of excitations in the system, and the dissipation operators in
(\ref{eq:evolution}) only decrease the number of excitations by dumping them
to the sink or to the reaction center. There are two different ways to answer
this question. The first one is simple yet somewhat formal---Vedral
established the relative entropy of entanglement in the context of quantum
information theory as a measure of entanglement for a general set of states
with no subspace limitations imposed. Therefore, even though the form of
specific Hamiltonian and dissipation terms in (\ref{eq:evolution}) leads to
the absence of multiply-excited states in the density operator, there is no
particular reason to consider that multiply-excited states and the mode of
preparation of a state are irrelevant in this setting. From this perspective,
entanglement is determined solely by the state rather than by the history and
mode of state evolution. The relative entropy of entanglement is a measure of
the \textquotedblleft distance\textquotedblright\ of a state to the closest
separable state and excluding multiply-excited states from the set of
separable states introduces a distortion to the original concept. Actually,
the measure of such a distortion may be rather large, as we found out
performing optimization of $D(\rho^{AB}||\sigma^{AB})$ in the full space (see
Figure~\ref{fig:dima-plots}(a)).%

Our second answer is more physically motivated. In general, while analyzing
the exciton dynamics in a light harvesting system, one assumes that the
intensity of light, which determines the photon flux, is quite small. Then it
follows that the probability of simultaneous creation of two excitations is
negligible. By performing the optimization of the $D(\rho^{AB}||\sigma^{AB})$
in the full space, which includes multiply-excited states that can have the
form $\left\vert E\right\rangle ^{A}\left\vert E\right\rangle ^{B}$, and
computing the amount of multiply-excited state population in the closest
separable state $\sigma^{AB}$, we can set up a limit for the light intensity
when the single-excitation assumption will become meaningless even in the
context of excitation-preserving dynamics. In accordance with Caratheodory's
theorem, we generated the generic separable state $\sigma^{AB}$ as a sum of
$2^{6}=64$ arbitrary pure separable states $\sum\limits_{n=1}^{64}%
{c_{n}\left\vert {\psi_{n}}\right\rangle }\left\langle {\psi_{n}}\right\vert
$. Here, we limited our computation to a simple example when site $A={1,6}$
and site $B={3}$, such that%
\[
\left\vert {\psi_{n}}\right\rangle =\left(  {\alpha_{o}\left\vert
{g}\right\rangle _{1}\left\vert {g}\right\rangle _{6}+\alpha_{1}\left\vert
{e}\right\rangle _{1}\left\vert {g}\right\rangle _{6}+\alpha_{2}\left\vert
{g}\right\rangle _{1}\left\vert {e}\right\rangle _{6}+\alpha_{3}\left\vert
{e}\right\rangle _{1}\left\vert {e}\right\rangle _{6}}\right)  \otimes\left(
{\beta_{1}\left\vert {g}\right\rangle _{3}+\beta_{2}\left\vert {e}%
\right\rangle _{3}}\right)  .
\]
The optimization is performed by the gradient method in the space of
coefficients $c_{n}$, $\alpha_{i}$, and $\beta_{j}$.
Figure~\ref{fig:dima-plots}(a) shows results of the computation for $T=77%
{{}^\circ}%
$K. We note that relative entropy of entanglement for the
single-excitation-subspace case has a similar shape to the full-space relative
entropy of entanglement, but it is approximately five times smaller.
Figure~\ref{fig:dima-plots}(b) demonstrates that the population of
doubly-excited states in the optimal separable state is negligibly small---it
is three to five orders of magnitude smaller than the population of
singly-excited states. We find it remarkable that such a negligible admixture
of doubly-excited states changes the relative entropy of entanglement by more
than a factor of five.

\section{Conclusions}

We presented results quantifying the quantum correlations present in a
biological system at cryogenic and physiological temperature.
Theorem~\ref{thm:main-theorem} proves that the single-excitation relative
entropy of entanglement admits a simple form. We then simulated the dynamics
of the FMO\ complex and calculated the quantum discord and the quantum mutual
information for various phenomenologically motivated bipartite cuts of the
sites in the FMO\ protein complex. It is surprising that the quantum discord
and the single-excitation relative entropy of entanglement are equivalent for
these simulations, but we have not been able to find a general proof that the
quantum discord of this system is equivalent to the formula in
Theorem~\ref{thm:main-theorem}. The results of our simulations indicate that
quantum correlations contribute a significant fraction of the total
correlation during the first picosecond of dynamics in many cases. Our last
contribution was to study the relative entropy of entanglement with an
unrestricted optimization, and we found that a small fraction of
doubly-excited states contribute significantly to reducing the relative
entropy of entanglement.

Open questions remain for this line of research. Our last contribution above
suggests an intriguing open question of whether optimizing the relative
entropy of entanglement over a single-excitation subspace, as done here and in
Ref.~\cite{SIFW09}, is operationally justified. We should also compute the
correlation of the quantum discord with the measure of transfer time in
Ref.~\cite{1367-2630-11-3-033003}\ of the excitation from the antenna to the
reaction center. This correlation might indicate how relevant
\textquotedblleft quantumness\textquotedblright\ is for the transfer of the
excitation. Finally, it might be interesting to determine how efficiency or
transfer time is affected by differing amounts of quantum discord, if there
can be efficient energy transfer without quantum discord, and how discord
compares with entanglement in the general case.

\begin{acknowledgments}
The authors acknowledge useful discussions with Mohan Sarovar and thank the
anonymous referees for useful comments. K.~B.~acknowledges support from the
Office of Naval Research under grant No. N000140811249. M.~M.~W.~acknowledges
support from the MDEIE (Qu\'{e}bec) PSR-SIIRI international collaboration
grant. D.~B.~U.~acknowledges support from the National Science Foundation
under grant No. PHY-0545390.
\end{acknowledgments}

\appendix

\section{}

\label{sec:app-proof}We first establish some notation before proving
Theorem~\ref{thm:main-theorem}. Let our zero- and single-excitation space be
spanned by the following states:%
\begin{align*}
\left\vert G\right\rangle ^{AB} &  \equiv\left\vert g\right\rangle
_{1}\left\vert g\right\rangle _{2}\cdots\left\vert g\right\rangle _{n},\\
\left\vert 1\right\rangle ^{AB} &  \equiv\left\vert e\right\rangle
_{1}\left\vert g\right\rangle _{2}\cdots\left\vert g\right\rangle _{n},\\
&  \vdots\\
\left\vert n\right\rangle ^{AB} &  \equiv\left\vert g\right\rangle
_{1}\left\vert g\right\rangle _{2}\cdots\left\vert e\right\rangle _{n}.
\end{align*}
The number of states is thus $n+1$. We introduce a bipartite splitting of the
above states so that the corresponding subsystems $A$ and $B$ are spanned by
$\{|G\rangle^{A},|1\rangle^{A},\dots,|n_{a}\rangle^{A}\}$ and $\{|G\rangle
^{B},\left\vert 1\right\rangle ^{B},\dots,|n_{b}\rangle^{B}\}$, respectively,
using the same convention as above. Note that $n_{a}+n_{b}+1=n$, and the
original basis written in terms of the split bases reads
\[
\{|G\rangle^{A}|G\rangle^{B},\ |k\rangle^{A}|G\rangle^{B},\ |G\rangle
^{A}|j\rangle^{B}\},
\]
where $k=1,\dots,n_{a}$ and $j=1,\dots,n_{b}$. Let $\Pi_{e}^{A}$ and $\Pi
_{e}^{B}$ denote the projectors onto the excited subspaces of the individual
subsystems of $A$ and $B$:%
\begin{equation}
\Pi_{e}^{A}\equiv\sum_{k=1}^{n_{a}}|k\rangle\langle k|^{A},\ \ \ \ \ \ \ \ \Pi
_{e}^{B}\equiv\sum_{j=1}^{n_{b}}|j\rangle\langle j|^{B}.
\end{equation}
The projector onto the full bipartite single-excitation subspace is as
follows:%
\begin{equation}
\Pi_{e}^{AB}\equiv|G\rangle\langle G|^{A}\otimes\Pi_{e}^{B}+\Pi_{e}^{A}%
\otimes|G\rangle\langle G|^{B}.\label{eq:single-excitation-projector}%
\end{equation}
Let $\Pi_{g}^{AB}\equiv|G\rangle\langle G|^{AB}$. Then $\Pi_{g}^{AB}+\Pi
_{e}^{AB}$ is a projector onto the zero- and single-excitation subspace. We
have that $\langle G|^{AB}\rho^{AB}|k\rangle^{AB}=0$ because the dynamics in
(\ref{eq:evolution}) do not induce any correlations between the ground state
and the single-excitation subspace of the density matrix $\rho^{AB}$. The
following chain of inequalities then gives another way to write the density
matrix $\rho^{AB}$ that is more useful to us:%
\begin{align}
\rho^{AB} &  =(\Pi_{g}^{AB}+\Pi_{e}^{AB})\rho^{AB}(\Pi_{g}^{AB}+\Pi_{e}%
^{AB})\nonumber\\
&  =\alpha\Pi_{g}^{AB}+\Pi_{e}^{AB}\rho^{AB}\Pi_{g}^{AB}+\Pi_{g}^{AB}\rho
^{AB}\Pi_{e}^{AB}+\Pi_{e}^{AB}\rho^{AB}\Pi_{e}^{AB}\nonumber\\
&  =\alpha\Pi_{g}^{AB}+\Pi_{e}^{AB}\rho^{AB}\Pi_{e}^{AB}\nonumber\\
&  =\alpha\Pi_{g}^{AB}+\left(  |G\rangle\langle G|^{A}\otimes\Pi_{e}^{B}%
+\Pi_{e}^{A}\otimes|G\rangle\langle G|^{B}\right)  \rho^{AB}\left(
|G\rangle\langle G|^{A}\otimes\Pi_{e}^{B}+\Pi_{e}^{A}\otimes|G\rangle\langle
G|^{B}\right)  \nonumber\\
&  =\alpha\Pi_{g}+\left(  |G\rangle\langle G|^{A}\otimes\Pi_{e}^{B}\right)
\rho^{AB}\left(  |G\rangle\langle G|^{A}\otimes\Pi_{e}^{B}\right)  +\left(
|G\rangle\langle G|^{A}\otimes\Pi_{e}^{B}\right)  \rho^{AB}\left(  \Pi_{e}%
^{A}\otimes|G\rangle\langle G|^{B}\right)  \nonumber\\
&  \ \ \ \ \ \ +\left(  \Pi_{e}^{A}\otimes|G\rangle\langle G|^{B}\right)
\rho^{AB}\left(  |G\rangle\langle G|^{A}\otimes\Pi_{e}^{B}\right)  +\left(
\Pi_{e}^{A}\otimes|G\rangle\langle G|^{B}\right)  \rho^{AB}\left(  \Pi_{e}%
^{A}\otimes|G\rangle\langle G|^{B}\right)  \nonumber\\
&  =\alpha\Pi_{g}+\rho_{e}^{A}\otimes|G\rangle\langle G|^{B}+|G\rangle\langle
G|^{A}\otimes\rho_{e}^{B}+\tau^{AB}.\label{eq:density-matrix-simplified}%
\end{align}
The first equality follows because $\rho^{AB}$ lives in the zero- and
single-excitation subspace. The second equality follows by expanding. The
third equality follows because the density operator $\rho^{AB}$ does not have
any correlations between the ground state $\left\vert G\right\rangle $ and any
excited state. The fourth equality follows from the relation in
(\ref{eq:single-excitation-projector}). The fifth equality follows by
expanding, and the last follows from the following definitions:%
\begin{align}
\rho_{e}^{A} &  \equiv\left(  \Pi_{e}^{A}\otimes|G\rangle\langle
G|^{B}\right)  \rho^{AB}\left(  \Pi_{e}^{A}\otimes|G\rangle\langle
G|^{B}\right)  ,\\
\rho_{e}^{B} &  \equiv\left(  |G\rangle\langle G|^{A}\otimes\Pi_{e}%
^{B}\right)  \rho^{AB}\left(  |G\rangle\langle G|^{A}\otimes\Pi_{e}%
^{B}\right)  ,\\
\tau^{AB} &  \equiv\left(  \Pi_{e}^{A}\otimes|G\rangle\langle G|^{B}\right)
\rho^{AB}\left(  |G\rangle\langle G|^{A}\otimes\Pi_{e}^{B}\right)  \\
&  \ \ \ \ \ \ +\left(  |G\rangle\langle G|^{A}\otimes\Pi_{e}^{B}\right)
\rho^{AB}\left(  \Pi_{e}^{A}\otimes|G\rangle\langle G|^{B}\right)  .\nonumber
\end{align}
The matrix $\rho_{e}^{A}$ is the projection of $\rho^{AB}$ into Alice's
single-excitation subspace, $\rho_{e}^{B}$ is similarly defined, and
$\tau^{AB}$ is the block-off-diagonal part of $\rho^{AB}$ living in the
single-excitation subspace.

\subsection{Proof that $R_{e}\left(  \rho^{AB}\right)  =H\left(
\overline{\Delta}\left(  \rho^{AB}\right)  \right)  -H\left(  \rho
^{AB}\right)  $}

\begin{proof}
Our first candidate to maximize the relative entropy of entanglement is a
state of the following form%
\begin{equation}
\sigma_{0}=\beta\Pi_{g}+\sigma_{e}^{A}\otimes|G\rangle\!\langle G|^{B}%
+|G\rangle\!\langle G|^{A}\otimes\sigma_{e}^{B}, \label{eq:candidatesep}%
\end{equation}
because it is in the most general form of a single-excitation separable state
with no correlation between the ground state and the other states. The
operators $\sigma_{e}^{A}$ and $\sigma_{e}^{B}$ are some positive operators
that live in the single-excitation subspaces with respective projectors
$\Pi_{e}^{A}$ and $\Pi_{e}^{B}$. So for now, we restrict the minimization in
the relative entropy of entanglement to states of the above form, leading to
the following minimization:%
\begin{equation}
\tilde{R}_{e}\left(  \rho^{AB}\right)  \equiv\min_{\sigma_{0}}\text{Tr}%
\left\{  \rho^{AB}\log\rho^{AB}\right\}  -\text{Tr}\left\{  \rho^{AB}%
\log\sigma_{0}\right\}  . \label{eq:rel-ent-almost}%
\end{equation}

We first determine a spectral decomposition of $\sigma_{e}^{A}$ and
$\sigma_{e}^{B}$ as follows:%
\begin{equation}
\sigma_{e}^{A}=\sum_{k=1}^{n_{a}}a_{k}|\phi_{k}\rangle\langle\phi_{k}%
|^{A},\ \ \ \ \ \ \ \sigma_{e}^{B}=\sum_{k=1}^{n_{b}}b_{k}|\phi_{k}%
\rangle\langle\phi_{k}|^{B}.
\end{equation}
Let $\omega^{A}\equiv\log\left(  {\sigma_{e}^{A}}\right)  $ and $\omega
^{B}\equiv\log\left(  {\sigma_{e}^{B}}\right)  $. The following equalities
then hold%
\begin{align*}
\tau^{AB}\log{\sigma_{0}} &  =\left[  \left(  \Pi_{e}^{A}\otimes
|G\rangle\langle G|^{B}\right)  \rho^{AB}\left(  |G\rangle\langle
G|^{A}\otimes\Pi_{e}^{B}\right)  +\left(  |G\rangle\langle G|^{A}\otimes
\Pi_{e}^{B}\right)  \rho^{AB}\left(  \Pi_{e}^{A}\otimes|G\rangle\langle
G|^{B}\right)  \right]  \\
&  \ \ \ \log\left(  \alpha\Pi_{g}+\sigma_{e}^{A}\otimes|G\rangle\!\langle
G|^{B}+|G\rangle\!\langle G|^{A}\otimes\sigma_{e}^{B}\right)  \\
&  =\left(  \Pi_{e}^{A}\otimes|G\rangle\!\langle G|^{B}\right)  \rho
^{AB}\left(  |G\rangle\!\langle G|^{A}\otimes\omega^{B}\right)  +\left(
|G\rangle\!\langle G|^{A}\otimes\Pi_{e}^{B}\right)  \rho^{AB}\left(
\omega^{A}\otimes|G\rangle\!\langle G|^{B}\right)
\end{align*}
The first equality follows by definition, and the second follows by
considering the overlapping support of the different subspaces. We then find
that%
\begin{align*}
\text{Tr}\left\{  \tau^{AB}\log{\sigma_{0}}\right\}   &  =\text{Tr}\left\{
\left(  \Pi_{e}^{A}\otimes|G\rangle\!\langle G|^{B}\right)  \rho^{AB}\left(
|G\rangle\!\langle G|^{A}\otimes\omega^{B}\right)  +\left(  |G\rangle\!\langle
G|^{A}\otimes\Pi_{e}^{B}\right)  \rho^{AB}\left(  \omega^{A}\otimes
|G\rangle\!\langle G|^{B}\right)  \right\}  \\
&  =\text{Tr}\left\{  \left(  |G\rangle\!\langle G|^{A}\otimes\omega
^{B}\right)  \left(  \Pi_{e}^{A}\otimes|G\rangle\!\langle G|^{B}\right)
\rho^{AB}\right\}  +\text{Tr}\left\{  \left(  \omega^{A}\otimes|G\rangle
\!\langle G|^{B}\right)  \left(  |G\rangle\!\langle G|^{A}\otimes\Pi_{e}%
^{B}\right)  \rho^{AB}\right\}  \\
&  =0.
\end{align*}
The second equality follows from cyclicity of the trace, and the last equality
follows because the zero- and single-excitation subspaces are orthogonal.
Therefore,%
\begin{equation}
\text{Tr}\left\{  \rho^{AB}\log{\sigma_{0}}\right\}  =\text{Tr}\left\{
(\rho^{AB}-\tau^{AB})\log{\sigma_{0}}\right\}  .
\end{equation}
Observe that $\rho^{AB}-\tau^{AB}$ is a state of the form $\sigma_{0}$ with
$\beta=\alpha$, $\sigma_{e}^{A}=\rho_{e}^{A}$, and $\sigma_{e}^{B}=\rho
_{e}^{B}$. Consider that%
\[
\text{Tr}\left\{  (\rho^{AB}-\tau^{AB})\log(\rho^{AB}-\tau^{AB})\right\}
\geq\text{Tr}\left\{  (\rho^{AB}-\tau^{AB})\log{\sigma_{0}}\right\}  ,
\]
because the relative entropy $D\left(  (\rho^{AB}-\tau^{AB})||\sigma
_{0}\right)  \geq0$~\cite{Nielsen:2000:CambridgeUniversityPress}. We conclude
that $\tilde{R}_{e}\left(  \rho^{AB}\right)  $ in (\ref{eq:rel-ent-almost})
attains its global minimum at $\rho-\tau^{AB}$. Thus,%
\[
\tilde{R}_{e}\left(  \rho^{AB}\right)  =H\left(  \overline{\Delta}\left(
\rho^{AB}\right)  \right)  -H\left(  \rho^{AB}\right)  .
\]

The expression in~(\ref{eq:candidatesep}) is not the most general
single-excitation separable state because it does not include coherence
elements between the ground state and the single-excitation subspace. That is,
matrix elements of the following form could be nonzero:%
\begin{align}
&  \left\vert G\right\rangle ^{A}\left\vert G\right\rangle ^{B}\left\langle
\phi_{k}\right\vert ^{A}\left\langle G\right\vert ^{B}+\text{H.C.}\\
&  \left\vert G\right\rangle ^{A}\left\vert G\right\rangle ^{B}\left\langle
G\right\vert ^{A}\left\langle \phi_{k}\right\vert ^{B}+\text{H.C.}%
\end{align}
We now show that it is not necessary to consider states of this more general
form by appealing to a perturbation theory argument \cite{K95}. Consider a
solution $\omega\equiv\rho-\tau^{AB}$ of (\ref{eq:rel-ent-almost}). Let us
take a small shift of this solution:%
\begin{equation}
\omega\rightarrow\omega+\epsilon\big(\left\vert G\right\rangle ^{A}\left\vert
G\right\rangle ^{B}\left\langle \phi_{k}\right\vert ^{A}\left\langle
G\right\vert ^{B}+\text{H.C.}\big).
\end{equation}
Consider that%
\begin{equation}
\omega|\phi_{k}\rangle^{A}|G\rangle^{B}=a_{k}|\phi_{k}\rangle^{A}|G\rangle
^{B},
\end{equation}
and%
\begin{equation}
\omega|G\rangle^{A}|G\rangle^{B}=\alpha|\phi_{k}\rangle^{A}|G\rangle^{B}.
\end{equation}

Since $\omega$ is Hermitian, we can apply perturbation theory to compute the
new states and energies. Brillouin-Wigner perturbation theory states that if
such a Schr\"{o}dinger-like equation above is perturbed via%
\begin{equation}
\omega\rightarrow\omega+\epsilon d\omega.
\end{equation}
where $d\omega=|G\rangle^{A}|G\rangle^{B}\langle\phi_{k}|^{A}\langle
G|^{B}+\ $H.C., then the new states are given by%
\[
|\phi_{k}\rangle^{A}|G\rangle^{B}\rightarrow|\phi_{k}\rangle^{A}|G\rangle
^{B}+\epsilon\sum_{|\phi_{m}\rangle\neq|\phi_{k}\rangle^{A}|G\rangle^{B}}%
|\phi_{m}\rangle\frac{1}{a_{k}-E_{m}}\langle\phi_{m}|d\omega|\phi_{k}%
\rangle^{A}|G\rangle^{B}+O(\epsilon^{2}).
\]
Here $E_{m}$ refers to the eigenvalue of the state $|\phi_{m}\rangle$.
Substituting for $d\omega$, we obtain%
\[
|\phi_{k}\rangle^{A}|G\rangle^{B}\rightarrow|\phi_{k}\rangle^{A}|G\rangle
^{B}+\frac{\epsilon}{a_{k}-\alpha}|G\rangle^{A}G\rangle^{B}+O(\epsilon^{2}).
\]
Similarly, applying BWPT to the second Schr\"{o}dinger like equation above
yields%
\[
|G\rangle^{A}|G\rangle^{B}\rightarrow|G\rangle^{A}|G\rangle^{B}+\epsilon
\sum_{|\phi_{m}\rangle\neq|G\rangle^{A}|G\rangle^{B}}|\phi_{m}\rangle\frac
{1}{\alpha-E_{m}}\langle\phi_{m}|d\omega|G\rangle^{A}|G\rangle^{B}%
+O(\epsilon^{2}).
\]
Again, substituting for $d\omega$ gives%
\[
|G\rangle^{A}|G\rangle^{B}\rightarrow|G\rangle^{A}|G\rangle^{B}+\frac
{\epsilon}{\alpha-a_{k}}|\phi_{k}\rangle^{A}G\rangle^{B}+O(\epsilon^{2}).
\]

Within BWPT, the eigenvalues are calculated as%
\[
a_{k}\rightarrow a_{k}+\epsilon\langle\phi_{k}|^{A}\langle G|^{B}d\omega
|\phi_{k}\rangle^{A}|G\rangle^{B}+\sum_{|\phi_{m}\rangle\neq|k\rangle
^{A}|G\rangle^{B}}\frac{\epsilon^{2}}{a_{k}-E_{m}}|\langle\phi_{m}%
|d\omega|\phi_{k}\rangle^{A}|G\rangle^{B}|^{2}%
\]
and similarly%
\[
\alpha\rightarrow\alpha+\epsilon\langle G|^{A}\langle G|^{B}d\omega
|G\rangle^{A}|G\rangle^{B}+\sum_{|\phi_{m}\rangle\neq|G\rangle^{A}%
|G\rangle^{B}}\frac{\epsilon^{2}}{\alpha-E_{m}}|\langle\phi_{m}|d\omega
|G\rangle^{A}|G\rangle^{B}|^{2}%
\]
Using the above, we find that the eigenvectors transform up to first order as
follows:%
\begin{align}
\left\vert \phi_{k}\right\rangle ^{A}\left\vert G\right\rangle ^{B}  &
\rightarrow\left\vert \phi_{k}\right\rangle ^{A}\left\vert G\right\rangle
^{B}+\epsilon{\frac{\left\vert G\right\rangle ^{A}\left\vert G\right\rangle
^{B}}{a_{k}-\alpha}}+O(\epsilon^{2}),\\
\left\vert G\right\rangle ^{A}\left\vert G\right\rangle ^{B}  &
\rightarrow\left\vert G\right\rangle ^{A}\left\vert G\right\rangle
^{B}+\epsilon{\frac{\left\vert \phi_{k}\right\rangle ^{A}\left\vert
G\right\rangle ^{B}}{\alpha-a_{k}}}+O(\epsilon^{2}).
\end{align}
Correspondingly, the eigenvalues change up to second order as follows:%
\begin{align}
a_{k}  &  \rightarrow a_{k}+O(\epsilon^{2}),\\
\alpha &  \rightarrow\alpha+O(\epsilon^{2}),
\end{align}
and so%
\[
\omega\rightarrow\omega+{\frac{\epsilon}{\alpha-a_{k}}}\log{\left(
\frac{\alpha}{a_{k}}\right)  }\Big[\left\vert G\right\rangle ^{A}\left\vert
G\right\rangle ^{B}\!\left\langle \phi_{k}\right\vert ^{A}\left\langle
G\right\vert ^{B}+\text{H.C.}\Big]+O(\epsilon^{2}).
\]
Using
\begin{equation}
\mathop{{\mathrm{Tr}}_{}}\big[\rho(\left\vert G\right\rangle ^{A}\left\vert
G\right\rangle ^{B}\!\left\langle \phi_{k}\right\vert ^{A}\left\langle
G\right\vert ^{B}+\text{H.C.})\big]=0
\end{equation}
we conclude that $\omega=\rho-\tau^{AB}$ is the state where the relative
entropy attains its global minimum.
\end{proof}

\bibliographystyle{unsrt}
\bibliography{Ref}

\end{document}